# KANEL: <u>K</u>olmogorov-<u>A</u>rnold <u>N</u>etwork <u>E</u>nsemble <u>L</u>earning Enables Early Hit Enrichment in High-Throughput Virtual Screening

Pavel Koptev[1], Nikita Krainov[1], Konstantin Malkov[1,2], Alexander Tropsha[2,3]

*1 AI1 Technologies*
*2 KAN Technologies, LLC*
*3 UNC-Chapel Hill*

## Abstract

Machine learning models of chemical bioactivity are increasingly used for prioritizing a small number of compounds in virtual screening libraries for experimental follow-up. In these applications, assessing model accuracy by early hit enrichment such as Positive Predicted Value (PPV) calculated for top N hits (PPV@N) is more appropriate and actionable than traditional global metrics such as AUC. We present KANEL, an ensemble workflow that combines interpretable Kolmogorov-Arnold Networks (KANs) with XGBoost, random forest, and multilayer perceptron models trained on complementary molecular representations (LillyMol descriptors, RDKit-derived descriptors, and Morgan fingerprints). Across five public PubChem BioAssay datasets (AIDs 485314, 485341, 504466, 624202, and 651820), Optuna-optimized weighted ensembles consistently outperformed the best single model in PPV@128 by 0.06–0.12 (9–40%). Models built with Morgan fingerprints substantially outperformed those built with LillyMol descriptors, while ensembles of specialty individual models outperformed single models trained on concatenated features. Y-randomization tests showed large performance degradation, supporting the notion that models built with unperturbed labels learned non-spurious structure-activity relationships. A preliminary graph neural network model for AID 504466 suggested the usefulness of this approach as a future ensemble component. Together, these results position KANEL as a reliable and interpretable workflow for early hit triage in data-driven drug discovery.

**Keywords:** virtual screening; QSAR; PPV@K; ensemble learning; Kolmogorov-Arnold networks; molecular descriptors; early hit enrichment

## 1 Introduction

Recent rapid expansion of fully enumerated chemical libraries[1] accessible for virtual screening (VS) has created substantial challenges for computational methods. For instance, Enamine REAL SPACE[2] currently counts over 78.1B make-on-demand molecules making the application of methods such as molecular docking practically impossible. Consequently, much more computationally efficient machine learning (ML) methods such as Deep Docking[3], HIDDEN GEM[4], and Deep QSAR[5] have been increasingly employed for virtual screening. However, in addition to efficiency requirements, the challenge of screening ultra-large chemical libraries also created a need to reconsider metrics for assessing ML model accuracy. Indeed, regardless of the size of VS libraries, only a small fraction of ranked compounds can be tested experimentally. In that regime, the key operational question is not whether a model ranks the entire library well on average as typically assessed by metrics such as AUC or Balanced Accuracy, but whether the very top of the ranked list is enriched in true actives. In our recent study[6], we argued that metrics such as PPV calculated for the top N ranked compounds (PPV@N) provide highly meaningful alternative to traditional ML model accuracy metrics. We further asserted and demonstrated that training set models built for highly imbalanced datasets and optimized for PPV afford much higher early enrichment in VS applications than models built with balanced datasets where the original excessive class of negatives is downsampled to match in size the smaller class of actives. We proposed using PPV@128 metric to compare and rank VS models because most experimental screening campaigns use plated sets of compounds with the number of wells in the plate proportional to 128, e.g., standard 384 well plates for screening 128 compounds in triplicates. We argued that metrics such as PPV@128 align more directly with realms of experimental medicinal chemistry triage than global ranking metrics.

In continuation of our previous work[6], herein we introduce KANEL as a modeling workflow developed to address the early-enrichment regime by combining diverse model families and feature sets in carefully validated ensembles. A distinguishing component is the use of Kolmogorov-Arnold Networks (KANs)[7], which are attractive because they offer interpretable learned univariate response functions while still contributing to ensemble diversity. Initial applications of KANs to molecular modeling have appeared in recent literature[8]. The goal of this pilot study is to present a concise summary of the current KANEL benchmark results, emphasizing methodological clarity, realistic claims, and practical implications for hit triage.

## 2 Related Work

Recent work has argued that enrichment-oriented metrics such as PPV@K[6] and BEDROC[9] are often more informative than global metrics for virtual screening workflows operating under strict experimental budgets. The benchmark framing adopted here follows that logic and uses public HTS assays to evaluate early hit enrichment under repeated train/test splits.

A related strand of work by members of the same broader development team comes from ai1 ScoreAI, a weighted-ensemble platform for lending and credit-risk assessment. In that setting, KAN components were combined with XGBoost and MLP models, and the published white paper reported 98.69% test accuracy (F1 = 0.99) for the weighted ensemble, with lender-specific tuning reaching 98.92% accuracy on an external loan-office dataset.[10] Although fintech classification and virtual screening are scientifically distinct tasks, this prior work is relevant as an engineering precedent: it supports the transferability of the team's ensemble-design, interpretability, and deployment-oriented optimization philosophy across high-stakes domains.

## 3 Methods

### 3.1 Benchmark datasets

We evaluate five public PubChem BioAssay datasets: AID 485314, 485341, 504466, 624202, and 651820. These were the same datasets using in our recent study[6], providing natural benchmark to assess relative performance of KANEL. All five datasets are strongly imbalanced, with active fractions ranging from 0.53% to 4.12%. Table 1 summarizes the dataset sizes and class imbalance.

*Table 1. Benchmark dataset sizes and class imbalance.*

| Dataset (AID) | Total size, molecules | Active molecules | Inactive molecules | Active (%) |
|---|---|---|---|---|
| 485314 | 315,598 | 4,430 | 311,168 | 1.40 |
| 485341 | 324,711 | 1,713 | 322,998 | 0.53 |
| 504466 | 299,377 | 4,081 | 295,296 | 1.36 |
| 624202 | 361,858 | 3,956 | 357,902 | 1.09 |
| 651820 | 277,632 | 11,451 | 266,181 | 4.12 |

### 3.2 Molecular representations

Three descriptor families were used in the main benchmark: LillyMol molecular descriptors, RDKit-derived molecular descriptors, and Morgan circular fingerprints. In the dedicated feature-engineering study, Morgan fingerprints were evaluated in a 2048-bit configuration with optimized settings. In the multi-model ensemble benchmark, a 256-bit Morgan fingerprint representation was used for computational efficiency. We also compared a single model trained on concatenated features against ensembles built from models specialized to individual feature sets.

### 3.3 Base models

The evaluated model families included XGBoost, random forest (RF), multilayer perceptron (MLP), and two KAN[7] variants: FasterKAN[11] and ReluKAN[12]. Because the best-performing single model varied across datasets and feature sets, KAN models should be viewed here as ensemble components that add diversity and interpretability potential rather than as a uniformly dominant single-model solution. For consistency of benchmarking against published data[6], since the published models did not consider ensembles, we developed single models using similar ML methods as in the previous publication[6] and reported our single model statistics in comparison with ensemble models.

### 3.4 Training and validation protocol

For each dataset, molecules were split into training (80%) and test (20%) sets five times using stratified sampling, i.e., such that each training and test set samples preserved the same ratio of actives and inactives as in the overall dataset. Within each split, hyperparameters were optimized with Optuna[13] using 5-fold cross-validation on the training subset. Fifty optimization trials were run with a Hyperband pruner, and the optimization objective was mean PPV@512 across the inner validation folds. This primary optimization metric was chosen after some experimentation that showed its positive impact on the performance of the earlier enrichment metrics such as PPV@128. After hyperparameter selection, the optimal model was retrained on the full training subset and evaluated once on the held-out test subset. Reported values are mean ± standard deviation across the five outer splits.

### 3.5 Ensemble construction and evaluation

We compared several fusion strategies, including element-wise product, arithmetic mean, and an Optuna-optimized weighted ensemble applied to predicted class probabilities. In the multi-representation setting, the weighted ensemble combines 12 base models: four model families trained separately on each of three feature sets. The primary external accuracy evaluation metric is PPV@128, which measures the fraction of true actives among the top 128 ranked molecules. We also report ROC-AUC, average precision / PR-AUC, and BEDROC at $\alpha = 20$ and $\alpha = 100$. Because PPV@128 depends on the test-set size and evaluation design, direct comparison with studies using different protocols requires caution.

## 4 Results

The central question in these experiments was whether carefully tuned ensembles deliver more actionable early enrichment than the best single model. As we show and discuss below, across the five assays considered here, the answer is resounding yes.

### 4.1 Ensemble performance across five assays

Table 2 compares the best single model with three ensemble rules across the five benchmark assays. Across all five datasets, the Optuna-weighted ensemble gives the highest mean PPV@128. Absolute gains over the best single model across all five datasets ranged from 0.06 to 0.12, corresponding to relative improvements of 9% to 40%. The largest absolute gain of 0.12 for a single dataset was for AID 624202 (improving PPV@128 from 0.36 to 0.48), while the largest relative gain was observed for AID 485341 (improving PPV@128 from 0.15 to 0.21, or 40%), which is also the most imbalanced dataset in the set.

*Table 2. PPV@128 (mean ± std over five splits) for ensemble strategies and the best single model across five datasets.*

| Dataset (AID) | Product | Arithmetic mean | Weighted | Best single model |
|---|---|---|---|---|
| 504466 | 0.83 ± 0.02 | 0.83 ± 0.02 | 0.88 ± 0.03 | 0.79 ± 0.04 |
| 485341 | 0.17 ± 0.02 | 0.16 ± 0.02 | 0.21 ± 0.02 | 0.15 ± 0.02 |

| Dataset (AID) | Product | Arithmetic mean | Weighted | Best single model |
|---|---|---|---|---|
| 485314 | 0.81 ± 0.02 | 0.82 ± 0.03 | 0.86 ± 0.02 | 0.77 ± 0.03 |
| 624202 | 0.40 ± 0.05 | 0.40 ± 0.03 | 0.48 ± 0.03 | 0.36 ± 0.05 |
| 651820 | 0.88 ± 0.02 | 0.89 ± 0.02 | 0.94 ± 0.02 | 0.86 ± 0.02 |

### 4.2 Multi-metric example on AID 504466

This dataset provides a representative example in which the weighted ensemble improves both early-enrichment and global ranking metrics. Table 3 shows gains in PPV@128, ROC-AUC, and BEDROC relative to the best single model. Figure 1 complements this table by showing the PPV-versus-TopK curve for the Optuna-weighted ensemble.

*Table 3. Multi-metric performance on AID 504466 (mean ± std over five splits).*

| Model | PPV@128 | ROC-AUC | BEDROC ($\alpha$ = 20) | BEDROC ($\alpha$ = 100) |
|---|---|---|---|---|
| Best single model (FasterKAN, RDKit) | 0.79 ± 0.04 | 0.91 ± 0.00 | 0.62 ± 0.01 | 0.52 ± 0.01 |
| Weighted ensemble (Optuna) | 0.88 ± 0.03 | 0.93 ± 0.00 | 0.67 ± 0.01 | 0.59 ± 0.01 |

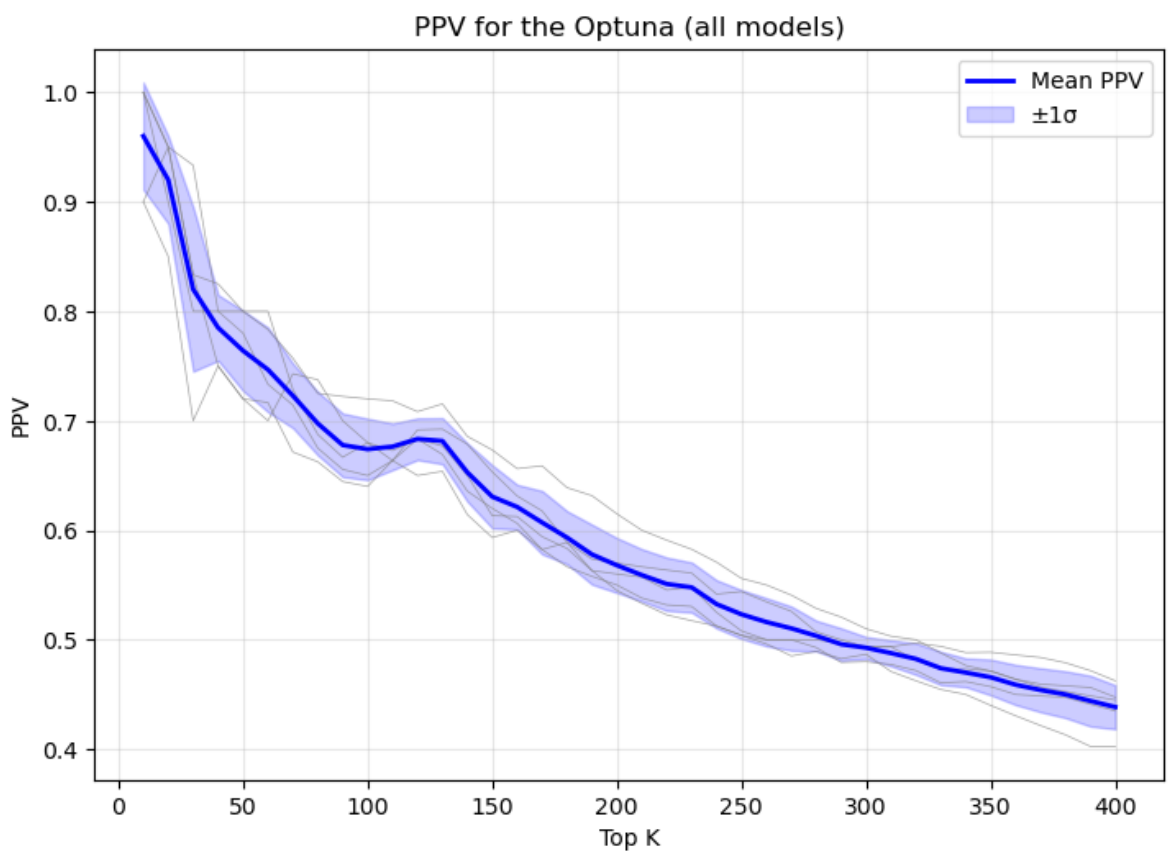

*Figure 1. PPV-versus-TopK curve for the Optuna-optimized weighted ensemble on AID 504466. The shaded band shows ±1 standard deviation across splits.*

### 4.3 Feature representation study

The feature-engineering study isolates representation effects using XGBoost across ten random splits. Morgan fingerprints clearly outperform LillyMol descriptors on all three reported metrics, and mixing Morgan fingerprints with LillyMol descriptors yields a further but modest improvement.

*Table 4. Feature representation comparison with XGBoost (mean ± std over 10 splits). Morgan results use the optimized 2048-bit fingerprint in the dedicated feature study.*

| Feature set | PR-AUC | PPV@128 | PPV@512 |
|---|---|---|---|
| LillyMol | 0.22 ± 0.01 | 0.45 ± 0.04 | 0.32 ± 0.01 |
| Morgan (2048-bit) | 0.45 ± 0.02 | 0.81 ± 0.02 | 0.52 ± 0.01 |
| Mixed (LillyMol + Morgan) | 0.48 ± 0.02 | 0.83 ± 0.02 | 0.55 ± 0.02 |

### 4.4 Specialized-model ensembles versus models built with concatenated features

A common question is whether one should concatenate all available feature sets into a single model or, instead, train specialized models and combine them at the prediction level. Table 5 shows that the latter strategy is favored: the weighted ensemble over specialized models reaches PPV@128 = 0.88 on AID 504466, compared with 0.83 for an XGBoost model trained on the concatenated feature set, which is still better than the best single model.

*Table 5. Controlled comparison between a single model on concatenated features and ensembles over specialized models on AID 504466.*

| Metric | Best single model | XGBoost (concatenated features) | Weighted ensemble |
|---|---|---|---|
| PPV@128 | 0.79 | 0.83 | 0.88 |

### 4.5 Validation via Y-randomization

Y-randomization was used to test whether performance might be driven by chance correlations or possible data leakage. Table 6 shows strong degradation after 50% label corruption across LillyMol, RDKit, and Morgan-fingerprint models. Figure 2 extends that analysis to the Optuna-weighted ensemble on AID 504466 and shows a monotonic decrease in PPV@128 as the corruption ratio increases.

*Table 6. Examples of performance degradation after 50% label corruption on AID 504466.*

| Feature set | ROC-AUC | Average precision | PPV@128 | BEDROC |
|---|---|---|---|---|
| LillyMol (original) | 0.91 | 0.27 | 0.65 | 0.55 |
| LillyMol (50% randomized) | 0.67 | 0.07 | 0.29 | 0.28 |
| RDKit (original) | 0.93 | 0.41 | 0.81 | 0.66 |
| RDKit (50% randomized) | 0.71 | 0.11 | 0.37 | 0.33 |
| Morgan FP (original) | 0.92 | 0.37 | 0.82 | 0.61 |
| Morgan FP (50% randomized) | 0.70 | 0.09 | 0.32 | 0.30 |

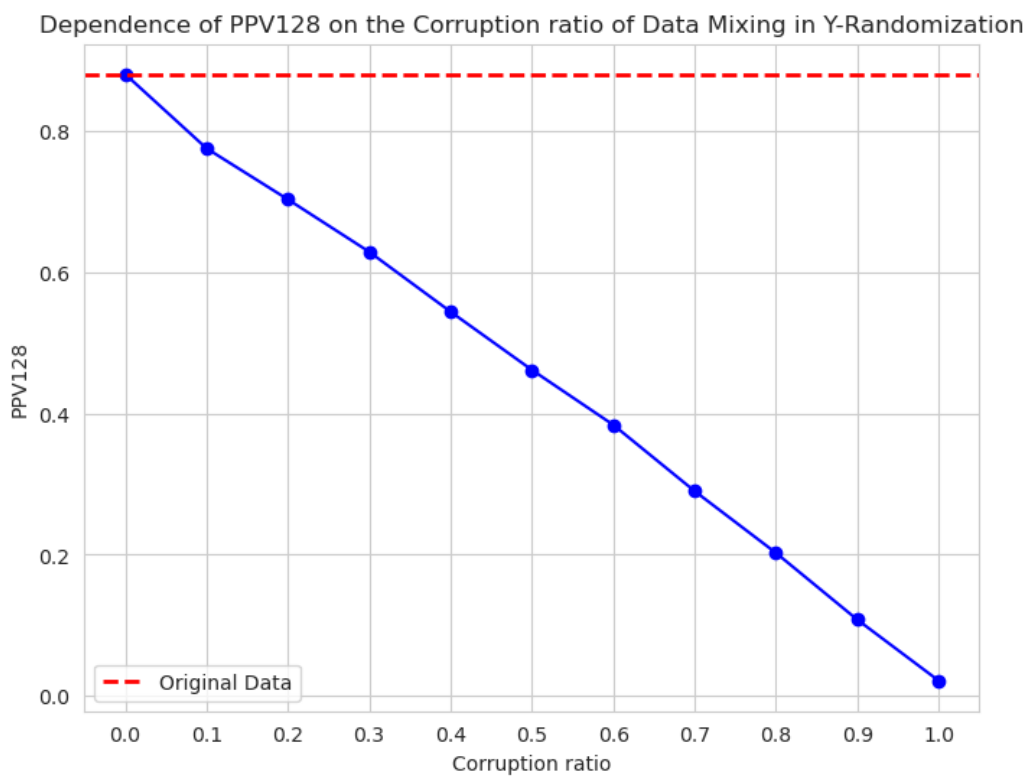

Figure 2. Dependence of ensemble PPV@128 on the corruption ratio in the Y-randomization test for AID 504466. Performance falls monotonically as the labels are progressively corrupted.

### 4.6 Preliminary studies using graph neural networks

A preliminary modeling study was carried out on AID 504466 using graph neural network (GNN) , molecular graph representations obtained with RDKit, and training a classical GNN on a single split. The resulting mean PPV@128 of 0.80 is competitive with the best descriptor-based single models, but it remains below the weighted ensemble. It is to early to assess this model's relative performance rigorously because the experiment was limited to one split and the computational cost is substantially higher compared to other methods. However, as GNN standalone model showed the best result by itself, while Ensemble model has shown overall better result, we plan to incorporate GNN as one of the ensemble models in future studies.

## 5 Discussion

Three points stand out as a result of this pilot study. First, early hit enrichment is the right optimization target for this application: the weighted ensembles deliver their clearest gains exactly where screening decisions are made, at the very top of the ranked list. Second, representation diversity matters. The gains do not arise from one universally superior base learner; they arise from combining complementary learners trained on complementary representations. Third, KAN models appear most valuable in the present study as interpretable ensemble components rather than as a standalone replacement for more established model classes.

Current benchmarking was performed with GPU-accelerated digital training. Emerging photonic and analog hardware platforms[14–16] may eventually offer attractive deployment paths for interpretable ensemble components, but no hardware acceleration results are claimed in the present study.

## 6 Limitations

(1) The study is based on repeated random stratified splits. Alternative evaluation protocols, especially scaffold-based splits, could produce different estimates of prospective generalization.

(2) PPV@128 is operationally meaningful but protocol-dependent. It is therefore useful for within-study model selection, but it should not be compared naively across studies that use different train/test designs or different effective TopK scales.

(3) Although KAN architectures are motivated in part by interpretability, this article does not yet include a formal interpretability analysis linking learned functions to chemically meaningful factors.

# 7 Conclusion

KANEL combines KAN variants, established machine-learning baselines, and complementary molecular representations in a validated ensemble workflow aimed at early hit enrichment. In this study, we focus on predictive performance; the interpretability of KAN-based models (and Ensembles containing KANs that proved high interpretability in real life ai1 Fintech Solutions) remains a discussion point, work in progress, and an important direction for future work. Across five public HTS datasets, Optuna-optimized weighted ensembles consistently outperform the best single model in PPV@128, with absolute gains of 0.06–0.12.

Feature studies show that representation choice matters strongly, with Morgan fingerprints outperforming LillyMol descriptors and prediction-level fusion outperforming simple feature concatenation. Y-randomization supports the view that the observed gains reflect learned structure-activity signal rather than accidental artifact.

The most important next steps are straightforward: extend the benchmark to additional assays, add scaffold-based evaluation, integrate additional models such as GNN predictions into the ensemble, quantify interpretability, and, most importantly, test prospective hit nomination in experimental drug discovery projects.